\documentclass[prd,showpacs,floatfix,twocolumn,,amsmath,amssymb,floatfix]{revtex4}
\usepackage{graphicx,color,dcolumn,booktabs,bm}
\usepackage{longtable,lscape}
\usepackage{txfonts}
\usepackage{overpic}
\usepackage{amssymb}
\usepackage{indentfirst}
\usepackage{feynmf}   
\usepackage{slashed}  
\usepackage{cases}
\usepackage{color}
\usepackage{multirow}
\usepackage{graphicx,color,dcolumn,booktabs,bm}

\graphicspath{{Figures/}} %

\graphicspath{{Figures/}} %
\usepackage[colorlinks, citecolor=black,anchorcolor=red,menucolor=red, linkcolor=red,filecolor=red,runcolor=red,urlcolor=blue,frenchlinks=red]{hyperref}
\usepackage{cases}

\begin{document}

\title{Phenomenological study of the isovector tensor meson family}
\author{Cheng-Qun Pang$^{1,2}$}{\email{pangchq13@lzu.edu.cn}
\author{Li-Ping He$^{1,2}$}\email{help08@lzu.edu.cn}
\author{Xiang Liu$^{1,2}$\footnote{Corresponding author}}\email{xiangliu@lzu.edu.cn}
\author{Takayuki Matsuki$^3$}\email{matsuki@tokyo-kasei.ac.jp}
\affiliation{$^1$School of Physical Science and Technology, Lanzhou University,
Lanzhou 730000, China\\
$^2$Research Center for Hadron and CSR Physics,
Lanzhou University $\&$ Institute of Modern Physics of CAS,
Lanzhou 730000, China\\
$^3$Tokyo Kasei University, 1-18-1 Kaga, Itabashi, Tokyo 173-8602, Japan}
\begin{abstract}

In this work, we study all the observed $a_2$ states and group them into the $a_2$ meson family, where their total and  two-body Okubo-Zweig-Iizuka allowed strong decay partial widths  are calculated via the quark pair creation model. Taking into account the present experimental data, we further give the corresponding phenomenological analysis, which is valuable to test whether each $a_2$ state can be assigned into the $a_2$ meson family. What is more important is that the prediction of their decay behaviors will be helpful for future experimental study of the $a_2$ states.

\end{abstract}

 \maketitle

\section{introduction}\label{sec1}

Among the observed light hadrons,  isovector tensor mesons form the $a_2$ meson family, which has the quantum number $I^GJ^{PC}=1^-2^{++}$. In Particle Data Group (PDG)  \cite{Beringer:1900zz}, seven $a_2$ states are collected, i.e., $a_2(1320)$, $a_2(1700)$, $a_2(1950)$, $a_2(1990)$, $a_2(2030)$, $a_2(2175)$, and  $a_2(2255)$. Here, their experimental information including the resonance parameters and the observed decay channels, is given in Table \ref{tab:1}.

\begin{table*}[htbp]
\caption{The resonance parameters and observed decay channels of the $a_2$ states collected in PDG \cite{Beringer:1900zz}. Here, the states listed as further states in PDG are marked by the superscript $\mathfrak{f}$.\label{tab:1}}
  \renewcommand{\arraystretch}{1.5}
 \tabcolsep=2pt
\begin{tabular}{lccccccc}
\toprule[1pt]
\toprule[1pt]
 {State} &                  {Mass (MeV)} &                    {Width (MeV)}       &Decay channel               \\     
 \midrule[1pt]
{$a_{2}(1320)$}&        {$1318.3^{+0.5}_{-0.6}$} &            {$105.0^{+1.6}_{-1.9}$}    &$\pi\rho$, $\pi \eta$, ${K K}$,
                                                                              $\pi\eta'$,\\&&&
                                                                             $\pi f_2(1270)$,
                                                                              $ \pi\rho (1450)$ \cite{Beringer:1900zz} \\

{$a_{2}(1700)$} &            {$1732\pm16$} &                   {$194\pm40$}  & $\pi \rho $, $\pi \eta$, $KK$,
                                                                              $\pi f_2(1270)$, $\rho \omega $ \cite{Beringer:1900zz} \\     

{$a_{2}(1950)^\mathfrak{f}$} &            {$1950^{+30}_{-70}$} &           {$180^{+30}_{-70}$}  & $\pi f_2(1270)$ \cite{Anisovich:2001pn}   \\        

{$a_{2}(1990)^\mathfrak{f}$}&             {$2050\pm10\pm40$} &             {$190\pm22\pm100$}     & $\pi\eta $, $\pi\eta '$   \cite{Anisovich:2001pn, Bugg:2004xu}   \\     

{$a_{2}(2030)^\mathfrak{f}$} &             {$2030\pm20$} &                    {$205\pm30$}       &  $\pi f_2(1270)$, $ \pi\eta$, $ \pi\eta '$ \cite{Anisovich:2001pn, Bugg:2004xu}  \\               

{$a_{2}(2175)^\mathfrak{f}$}&            {$2175\pm40$} &                      {$310^{+90}_{-45}$}    &   $\pi f_2(1270)$ \cite{Anisovich:2001pn, Bugg:2004xu}\\          
{$a_{2}(2255)^\mathfrak{f}$} &            {$2255\pm20$} &                   {$230\pm15$}  &            $\pi f_2(1950)$, $\pi\eta$
 \cite{Anisovich:2001pn, Bugg:2004xu}     \\     
\bottomrule[1pt]
\bottomrule[1pt]
\end{tabular}
\end{table*}

$a_2(1320)$ can be well established to be the $1^3P_2$ state \cite{Caso:1998tx,Burakovsky:1999ug} which is the ground state of the $a_2$ meson family, while $a_2(1700)$ is the first radial
excitation of $a_2(1320)$  \cite{Anderson:1990mv,Ackleh:1991dy,Munz:1996hb,Barnes:1996ff,Amsler:2008zzb}.
As shown in Table \ref{tab:1}, there are five $a_2$ states listed as further states in PDG. $a_2(1950)$ was observed in the $\pi  f_2(1270)$ decay channel by SPEC \cite{Anisovich:2001pn} when fitting it with the Crystal
Barrel data. 
In addition, $a_2(1950)$ was also observed in the process $p \overline{p}\to \eta \eta \pi$ \cite{Anisovich:2001pp}.
Anisovich {\it et al}. indicated that there exists $a_2(1990)$ [or named $a_2(1980)$] in the reactions $p\overline{p}\to \pi \eta$, $p \overline{p}\to  \pi\eta'$ \cite{Anisovich:1999jv}. Furthermore, in
an analysis combined with the consistent resonance parameters in all three sets of data, ($p \overline{p}\to \pi \eta$, $\pi \eta' $, $3\pi$), the authors in Ref. \cite{Anisovich:2001pn} updated their analyzed results, in which
the former resonances $a_2(1990)$, $a_2(2270) $ \cite{Anisovich:1999jv}, $a_2(2100)$, and $a_2(2280)$ \cite{Anisovich:1999jv} are replaced with $a_2(1950)$, $a_2(2030)$, $a_2(2175)$, and $a_2(2255)$. Thus, in the 2002 PDG edition \cite{Hagiwara:2002fs}, $a_2(1990)$ was still listed while the updated states $a_2(1950)$,~$a_2(2030)$,~$a_2(2175)$, and $a_2(2255)$ were also included \cite{Masjuan:2012gc} . In Ref. \cite{Shchegelsky:2006es}, the $a_2$ state with resonance parameters, the mass {$M=2050\pm10\pm40$} MeV and width {$\Gamma=190\pm22\pm100$} MeV, was reported in the process $\gamma \gamma \to \pi ^+ \pi^- \pi^0$, where this $a_2$ is considered as the same state as $a_2(2030)$. In addition, the observed $a_2$ resonance with the mass {$2003\pm10\pm19$} MeV and width {$249\pm23\pm30$} MeV is treated as $a_2(1990)$ since its resonance parameters are close to those of $a_2(1990)$ \cite{Lu:2004yn}.

Although many $a_2$ states were reported by experiments, we notice that these states are not established, especially for the states listed as further states in PDG, which is the main reason why we are interested in the study of the $a_2$ states. It is obvious that the detailed information on the partial decay widths of the $a_2$ states is helpful for further experimental study on these. Comparing our theoretical results with the measured resonance parameters, we can further test whether they are suitably categorized into the $a_2$ meson family. In the next section, we briefly review the possible assignments of the states in the $a_2$ meson family. Considering the present research status of these states, we notice that
a systematic and phenomenological study of the $a_2$ is still absent. Hence, in this work we carry out the calculation of the Okubo-Zweig-Iizuka (OZI) allowed partial decay widths of these states, where the quark pair creation (QPC) model proposed by Micu \cite{Micu:1968mk} will be applied to the whole calculation. The systematical study will give us the valuable partial and total decay widths of
the discussed $a_2$ states in detail.

This paper is organized as follows. In the next section, we briefly introduce different categorizations of the states in the $a_2$ meson family. Further, we calculate the corresponding two-body OZI-allowed strong decays. After combining the present experimental data with our theoretical results, a phenomenological analysis will be given. The final section is devoted to a summary of our work.

\begin{table*}[htbp]
 \tabcolsep=12pt
  \renewcommand{\arraystretch}{1.5}
\caption{Different categorizations of the $a_2$ meson family. Here, the states with the superscript $\it{p}$ are the predicted states in the corresponding analysis.\label{tab:2}}
\begin{tabular}{ccccc}
\toprule[1pt]
\toprule[1pt]
$n^{2S+1}L_J$&  A. Anisovich \cite{Anisovich:2001pn,Bugg:2004xu,Bugg:2012yt} &  {V.~Anisovich}~\cite{Anisovich:2002us,Anisovich:2003tm}&Masjuan \cite{Masjuan:2012gc}\\ \midrule[1pt]
$1^3P_2$&$a_2(1320)$&$a_2(1320)$&$a_2(1320)$\\
$2^3P_2$&$a_2(1700)$&$a_2(1700)$&$a_2(1700)$\\
$3^3P_2$&$a_2(1950)$&$a_2(1950)$&$a_2(2175)$\\
$4^3P_2$&$a_2(2175)$&$a_2(2225)$&$a_2(2420)^p$\\
$1^3F_2$&$a_2(2030)$&$a_2(2030)$&$a_2(2030)$\\
$2^3F_2$&$a_2(2225)$&$a_2(2310)^p$&$a_2(2225)$\\
\bottomrule[1pt]
\bottomrule[1pt]
\end{tabular}
\end{table*}

\section{Mass spectrum analysis and calculation of two-body strong decays}\label{sec2}

\subsection{Mass spectrum analysis of the $a_2$ states}

\begin{center}
\begin{table}[htbp]
 \renewcommand{\arraystretch}{1.2}
 \tabcolsep=1.2pt
\caption{The OZI-allowed two-body decay channels of the discussed $a_2$ states. Here, $\rho$, $\omega$, $\eta$ and $\eta'$ denote $\rho(770)$, $\omega(782)$,~$\eta(548)$, and $\eta'(958)$,
respectively. We use $\surd$ to mark the allowed decay channels. \label{tab:3}}
\begin{tabular}{ccccccccc}
\toprule[1pt]
\toprule[1pt]
 {channel} &  \footnotesize{$a_{2}(1320)$} & \footnotesize{$a_{2}(1700)$} & \footnotesize{$a_{2}(1950)$} & \footnotesize{$a_{2}(2030)$} &  \footnotesize{$a_{2}(2175)$} &      \footnotesize{$a_{2}(2255)$}   \\     
 \midrule[1pt]
$\pi \eta$&$\surd$     &      $\surd$     &      $\surd$     &      $\surd$     &      $\surd$     &      $\surd$     \\
$\pi  \rho $     &      $\surd$     &      $\surd$     &      $\surd$     &      $\surd$     &      $\surd$     &      $\surd$    \\
$KK$     &      $\surd$     &      $\surd$     &      $\surd$     &      $\surd$     &      $\surd$     &      $\surd$      \\
$ \pi\eta ' $     &      $\surd$     &      $\surd$     &      $\surd$     &      $\surd$     &      $\surd$     &      $\surd$      \\
$\pi b_{1}(1235) $     &           &      $\surd$     &      $\surd$     &      $\surd$     &      $\surd$     &      $\surd$     \\
$KK^*$     &           &      $\surd$     &      $\surd$     &      $\surd$     &      $\surd$     &      $\surd$     \\
$\pi f_2(1270)$     &           &      $\surd$     &      $\surd$     &      $\surd$     &      $\surd$     &      $\surd$    \\
$\pi  f_1{}(1280)$     &           &      $\surd$     &      $\surd$     &      $\surd$     &      $\surd$     &      $\surd$     \\
$\pi \eta (1295)  $     &           &      $\surd$     &      $\surd$     &      $\surd$     &      $\surd$     &      $\surd$    \\
$\rho \omega$     &           &      $\surd$     &      $\surd$     &      $\surd$     &      $\surd$     &      $\surd$     \\
$\pi f_1{}(1420)$     &           &      $\surd$     &      $\surd$     &      $\surd$     &      $\surd$     &      $\surd$      \\
$\pi  \rho (1450)$     &           &      $\surd$     &      $\surd$     &      $\surd$     &      $\surd$     &      $\surd$      \\
$ \pi \eta (1475)$     &           &      $\surd$     &      $\surd$     &      $\surd$     &      $\surd$     &      $\surd$     \\
$\pi  f'_2(1525)$     &           &      $\surd$     &      $\surd$     &      $\surd$     &      $\surd$     &      $\surd$     \\
$\pi \eta _2{}(1645)$     &           &           &      $\surd$     &      $\surd$     &      $\surd$     &      $\surd$    \\
$K{K_1{}}(1270)$     &           &           &      $\surd$     &      $\surd$     &      $\surd$     &      $\surd$  \\
$\eta  a_1{}(1260)$     &           &           &      $\surd$     &      $\surd$     &      $\surd$     &      $\surd$    \\
$K^* K^*$     &           &           &      $\surd$     &      $\surd$     &      $\surd$     &      $\surd$     \\
$ \pi \rho _3{}(1690)$     &           &           &      $\surd$     &      $\surd$     &      $\surd$     &      $\surd$  \\
$\pi \eta (1300)$     &           &           &      $\surd$     &      $\surd$     &      $\surd$     &      $\surd$     \\
$\pi  \rho (1700)$     &           &           &      $\surd$     &      $\surd$     &      $\surd$     &      $\surd$   \\
$\eta  a_2{}(1320)$     &           &           &      $\surd$     &      $\surd$     &      $\surd$     &      $\surd$  \\
$K{K_1{}}(1400)$     &           &           &      $\surd$     &      $\surd$     &      $\surd$     &      $\surd$  \\
$K{K^*}(1410)$     &           &           &      $\surd$     &      $\surd$     &      $\surd$     &      $\surd$   \\
$K{K_2{}^*{}}(1430)$     &           &           &      $\surd$     &      $\surd$     &      $\surd$     &      $\surd$ \\
$\rho  h_1{}(1170)$     &           &           &      $\surd$     &      $\surd$     &      $\surd$     &      $\surd$ \\
$ \rho a_1{}(1260) $     &           &           &           &      $\surd$     &      $\surd$     &      $\surd$   \\
$ \omega b_1{}(1230) $     &           &           &           &      $\surd$     &      $\surd$     &      $\surd$   \\
$\pi  \rho (1900)$     &           &           &           &      $\surd$     &       $\surd$     &      $\surd$  \\
$ \rho \pi (1300) $     &           &           &           &           &    $\surd$     &      $\surd$\\
$ \rho a_2{}(1320) $     &           &           &           &           &       $\surd$     &      $\surd$\\
$\pi f_2{}(2010)$     &           &           &           &           &       $\surd$     &      $\surd$\\
$\pi f_4(2050)$     &           &           &           &           &          $\surd$     &      $\surd$\\
$ {K^*} K_1{}(1270)$     &           &           &           &           &            $\surd$     &      $\surd$\\
$\eta'a_1(1260)$     &           &           &           &           &                 &      $\surd$\\
$\rho  \omega (1420)$     &           &           &           &           &                  &      $\surd$\\
$KK^*(1680)$     &           &           &           &           &                 &      $\surd$\\
$\eta  \pi _2{}(1670)$     &           &           &           &             &           &      $\surd$\\
$\omega \rho (1450) $     &           &           &           &                     &           &      $\surd$\\
$ \rho a_0{}(1450)$     &           &           &           &                     &           &      $\surd$\\\bottomrule[1pt]
\bottomrule[1pt]
\end{tabular}
\end{table}
\end{center}

Before calculating the OZI-allowed two-body strong decay widths, we briefly review the status of the mass spectrum analysis of the $a_2$ states. Usually, the analysis of the Regge trajectories can be an effective approach to study the categorization of mesons.
In Refs. \cite{Anisovich:1999jx,Anisovich:2000kxa,Anisovich:2001pn}, Anisovich {\it et al.} studied the light meson spectrum via the analysis of the Regge trajectories. As indicated in Ref. \cite{Anisovich:2001pn}, $a_2(1320)$, $a_2(1700)$, $a_2(1950)$, and $a_2(2175)$ can be assigned to $1^3P_2$, $2^3P_2$, $3^3P_2$, and $4^3P_2$ states, respectively, while $a_2(2030)$ and $a_2(2255)$ are good candidates for the $1^3F_2$ and $2^3F_2$ states, respectively \cite{Anisovich:2001pn, Bugg:2004xu,Bugg:2012yt}. However, in Refs. \cite{Anisovich:2002us,Anisovich:2003tm} $a_2(2175)$ was not included in their analysis. Alternatively, $a_2(2255)$ is treated as the $4^3P_2$ state and $a_2(2310)$ as the $2^3F_2$ state \cite{Anisovich:2003tm}. Recently, Masjuan {\it et al.} \cite{Masjuan:2012gc} pointed out
that $a_2(1950)$ and $a_2(2030)$ might be the same state. Their analysis indicates that $a_2(1320)$, $a_2(1700)$, and $a_2(2175)$ are $1^3P_2$, $2^3P_2$ and $3^3P_2$ states, respectively. Furthermore, the $4^3P_2$ state with the mass 2.42(17) GeV was predicted in their trajectory  analysis. The obtained assignments to $a_2(2030)$ and $a_2(2255)$ in Ref. \cite{Masjuan:2012gc} are consistent with those in Ref. \cite{Anisovich:2001pn}. In Table \ref{tab:2}, we summarize the three different categorizations mentioned above.

\subsection{Brief review of the QPC model}
In this work, we study the two-body OZI-allowed strong decays of the $a_2$ states under the three categorizations. In the following, we briefly explain the QPC model adopted here. After the
QPC model \cite{Micu:1968mk} was proposed, this model was further developed by the Orsay group of Le Yaouanc {\it et~al.} \cite{Le Yaouanc:1972ae,Le Yaouanc:1973xz,Le Yaouanc:1974mr,Le Yaouanc:1977ux,LeYaouanc:1977gm}.
 Later, this model was widely applied  to study the OZI-allowed strong decay of hadrons
 \cite{vanBeveren:1982qb,Capstick:1993kb,Blundell:1995ev,Ackleh:1996yt,Capstick:1996ib,Close:2005se,Zhang:2006yj,Lu:2006ry,Sun:2009tg,Liu:2009fe,
Sun:2010pg,Rijken:2010zza,Yu:2011ta,Zhou:2011sp,Wang:2012wa,Ye:2012gu,Sun:2013qca,He:2013ttg,Sun:2014wea}.

For a two-body strong decay process $A\rightarrow B+C$, the corresponding transition matrix element can be written as
\begin{eqnarray}
\langle BC|T|A \rangle = \delta ^3(\mathbf{P}_B+\mathbf{P}_C)\mathcal{M}^{{M}_{J_{A}}M_{J_{B}}M_{J_{C}}},
\end{eqnarray}
where $\mathbf{P}_{B(C)}$ denotes the three-momentum of the final particle $B(C)$. $M_{J_i}$ ($i=A,\,B,\,C$) is the orbital magnetic momentum of the corresponding meson in the decay. $\mathcal{M}^{{M}_{J_{A}}M_{J_{B}}M_{J_{C}}}$ is the helicity amplitude. The $T$ operator reads as
\begin{eqnarray}
T& = &-3\gamma \sum_{m}\langle 1m;1~-m|00\rangle\int d \mathbf{p}_3d\mathbf{p}_4\delta ^3 (\mathbf{p}_3+\mathbf{p}_4) \nonumber \\
 && \times \mathcal{Y}_{1m}\left(\frac{\textbf{p}_3-\mathbf{p}_4}{2}\right)\chi _{1,-m}^{34}\phi _{0}^{34}
\omega_{0}^{34}b_{3i}^{\dag}(\mathbf{p}_3)d_{4j}^{\dag}(\mathbf{p}_4),
\end{eqnarray}
where $\gamma$ is a parameter that takes the value 8.7 or ${8.7}/{\sqrt{3}}$ depending on whether the quark-antiquark pair created from the vacuum
is $u\bar u(d\bar d)$ or $s\bar s$ \cite{Ye:2012gu}. The quark and antiquark created from the vacuum are marked by the subscripts 3 and 4, respectively. $i/j$ denotes the color indices. The $\chi$, $\phi$, and $\omega$  are the spin, flavor, and color wave functions, respectively.
In addition, $\mathcal{Y}_{lm}(\mathbf{p})=|\mathbf{p}|Y_{lm}(\mathbf{p}) $ is the  solid harmonic polynomial
(see Refs. \cite{Roberts:1992js,Blundell:1996as} for more details). Using the Jacob-Wick formula \cite{Jacob:1959at}, the helicity amplitude $\mathcal{M}^{{M}_{J_{A}}M_{J_{B}}M_{J_{C}}}$
can be converted into the partial wave amplitude $M^{JL}(\mathbf{P})$, i.e.,
\begin{eqnarray}
\mathcal{M}^{JL}(\mathbf{P})&=&\frac{\sqrt{2L+1}}{2J_A+1}\sum_{M_{J_B}M_{J_C}}\langle L0;JM_{J_A}|J_AM_{J_A}\rangle \nonumber \\
&&\times \langle J_BM_{J_B};J_CM_{J_C}|{J_A}M_{J_A}\rangle \mathcal{M}^{M_{J_{A}}M_{J_B}M_{J_C}}.
\end{eqnarray}
Finally, the decay width can be given by
\begin{eqnarray}
\Gamma&=&\pi ^2\frac{|\mathbf{P}|}{m_A^2}\sum_{J,L}|\mathcal{M}^{JL}(\mathbf{P})|^2,
\end{eqnarray}
where $m_A$ is the mass of the initial meson $A$.
In the concrete calculation, we use the harmonic oscillator wave function to describe the meson spacial wave function, where we approximately take the harmonic oscillator potential to describe the potential between quark and antiquark \cite{vanBeveren:1982sj}. The harmonic oscillator
wave function has the following expression: 
\begin{eqnarray}
\Psi_{nlm}(R,\mathbf{p})= \mathcal{R}_{nl}(R,\mathbf{p}) \mathcal{Y}_{lm}(\mathbf{p}),
\end{eqnarray}
where $R$ is a parameter, which is given in Ref.~\cite{Close:2005se} for the mesons involved in our calculation.

The two-body OZI-allowed strong decay channels, which are allowed by the conservation law and calculated by us, are listed in Table \ref{tab:3} for $a_2(1320)$, $a_2(1700)$, $a_2(1950)$, $a_2(2175)$, $a_2(2030)$, and $a_2(2255)$.

\subsection{Phenomenological analysis of two-body decays}

\subsubsection{$a_2(1320)$, $a_2(1700)$, and $a_2(2030)$}

From Table \ref{tab:2}, we notice that different groups obtained the consistent conclusion of the assignments to $a_2(1320)$, $a_2(1700)$, and $a_2(2030)$, which are the $1^3P_2$, $2^3P_2$, and $1^3F_2$ states, respectively. In the following, we discuss the decay behaviors of $a_2(1320)$, $a_2(1700)$, and $a_2(2030)$.

As for $a_2(1320)$, there are four allowed decay channels. We present the dependence of total and partial decay widths on the $R$ value in Fig. \ref{fig:1320}, and compare the obtained total decay width with the experimental data. The calculated total decay width of $a_2(1320)$ is consistent with the experimental data measured by Ref. \cite{Thompson:1997bs} when taking $R=3.85$ GeV$^{-1}$ ~\cite{Close:2005se}. Our result also shows that $\pi\rho$ is the dominant decay mode and $\pi\eta$ is another main decay mode.
Since there is the abundant experimental information on $a_2(1320)$, we  list the comparison between theoretical and experimental results of five typical ratios and two partial decay widths in Table \ref{tab:1320}, which indicates that our calculation is comparable with the present experimental data. Thus, $a_2(1320)$ as the $1^3P_2$ state is well tested by our study. 

From PDG \cite{Beringer:1900zz}, we also notice there were the measurement results of $a_2(1320) \to \omega \pi \pi$ and $a_2(1320) \to 3\pi$, where the breaching ratios of $a_2(1320) \to \omega \pi \pi$ and $a_2(1320) \to 3\pi$ can reach up to $(70.1\pm2.7)\%$ and $(10.6\pm3.2)\%$, respectively. This information also stimulates our interest in investigating these three-body decays. The two-body decay behavior of $a_2(1320)$ indicates that the $\pi\rho$ channel is its dominant decay mode, where $\rho$ dominantly decays into $2\pi$ [branching ratios ${\mathcal{B}}(\rho\to \pi\pi)\sim 100\%$]. In addition, the $2\pi$ in $a_2(1320)\to \omega \pi\pi$ can be from the intermediate $\rho$. Thus, we consider $a_2(1320)\to \pi\rho\to 3\pi$ and $a_2(1320)\to \omega\rho\omega \pi\pi$ processes, where the decay amplitudes of $a_2(1320)\to \pi\rho$ and $a_2(1320)\to \omega\rho$ can be obtained by the QPC model. And then we apply the general expressions for three-body decays, i.e.,
\begin{align}\label{equ_phase}
     \Gamma_{A\to B+C\to 1+2+C}=\int_{m_{1}+m_2}^{m_A-m_3}\frac{dE}{2\pi}\frac{ \Gamma_{A\to B+C} \, \Gamma_{B\to 1+2}}{(E-m_B)^2+\Gamma{_B}^2/4},
\end{align}
where $\Gamma _{B}$ is the  total width of meson $B$ and $E$ is the energy of particle $B$. By the calculation, 
finally we obtain $\Gamma_{a_2(1320)\to 3\pi} \simeq$  90 MeV and  $\Gamma_{a_2(1320)\to \omega\pi^+\pi^{-}} \simeq$ 15 MeV, which are comparable with the corresponding experimental data \cite{Beringer:1900zz}.

In addition, $a_2(1320)\to \pi^{\pm}\gamma$ can from $a_2(1320)\to \pi^{\pm}\rho\to \pi^{\pm}\gamma$ when one considers the Vector-Meson-Dominance mechanism. Since in this work we only focus on the strong decays, we do not discuss the $a_2(1320)\to \pi^{\pm}\gamma$ radiative decay.

\begin{figure}[htb]
\begin{center}
\scalebox{1}{\includegraphics[width=\columnwidth]{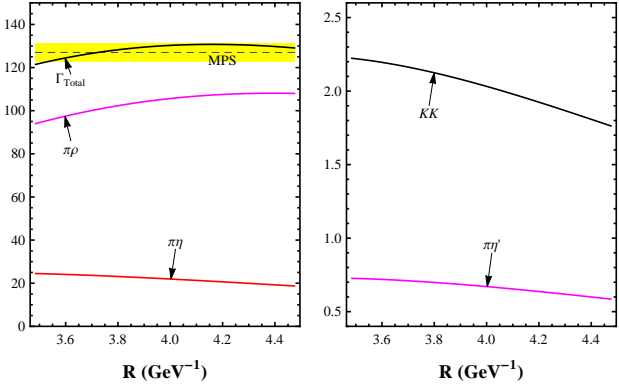}}
\caption{The partial and total decay widths of $a_2(1320)$ as the $1^3P_2$ state dependent on the $R$ value. Here, the dashed line with the yellow band is the experimental total width from Ref. \cite{Thompson:1997bs}.  All results are in units of MeV.} \label{fig:1320}
\end{center}
\end{figure}

\renewcommand{\arraystretch}{1.5}
\begin{table}[htbp]
\caption{The comparison between the calculated results and the experimental data for $a_2(1320)$. $\Gamma_{\pi\eta}$ and $\Gamma_{KK}$ are in units of MeV.
 \label{tab:1320}}
\begin{tabular}{lcccccccc}
\toprule[1pt]
\toprule[1pt]
&                                                  {This work} &                    {Experimental data}  \\     
 \midrule[1pt]
{$\Gamma _{ \pi\eta}/\Gamma _{Total}$} &                       $0.18$ &            {$(0.15 \pm 0.04)$}~\cite{Barnham:1971nw}\ \\     

{$\Gamma _{K K}/\Gamma _{Total}$} &                          {$0.016$} &              {$(0.049 \pm 0.008)$~\cite{Beringer:1900zz}}   \\     

{$\Gamma _{\pi\eta'}/\Gamma _{Total}$} &                    {$5.4\times 10^{-3}$} &        {$(5.3\pm 0.9)\times 10^{-3}$~\cite{Beringer:1900zz}}  \\     

{$\Gamma _{K K}/\Gamma _{\pi \eta}$} &                          {$0.092$} &              {$0.08\pm 0.02$}~\cite{Beringer:1900zz,Bertin:1998sb}   \\     

{$\Gamma _{\pi\eta'}/\Gamma _{\pi \eta}$} &                          {$0.030$} &              {$0.032\pm 0.009$}~\cite{Beringer:1900zz,Abele:1997dz}   \\     

{$\Gamma _{\pi \eta}$} &                                     {$23$} &                    {$18.5\pm 3.0$}~\cite{Beringer:1900zz,Shchegelsky:2006et} \\     

{$\Gamma _{K K}$} &                                            {$2.1$} &        {$7.0^{+2.0}_{-1.5}$}~\cite{Beringer:1900zz,Shchegelsky:2006et}   \\     

\bottomrule[1pt]
\bottomrule[1pt]
\end{tabular}
\end{table}
As for $a_2(1700)$, more decay channels open as shown in Table \ref{tab:2}. In Fig. \ref{fig:1700} we list the variation of the total and partial decay widths of $a_2(1700)$ in terms of the $R$ value. When $R=4.35$~GeV$^{-1}$~\cite{Close:2005se}, the central value of the experimental total width of $a_2(1700)$ from L3 \cite{Acciarri:2000ex} can be reproduced by our calculation. The main decay modes of $a_2(1700)$ include $\pi\rho$, $\rho\omega$, $\pi\eta$, and $\pi b_1(1235)$. Additionally,  $\pi f_1(1285)$, $\pi\eta^\prime$, $\pi\eta^\prime(1295)$, $\pi\rho(1450)$, and $KK$ are important decay channels for $a_2(1700)$. In Table \ref{tab:1700}, further theoretical values of $\Gamma_{Total}$, $\Gamma_{\pi\eta}$, $\Gamma_{KK}$ and $\Gamma_{\pi\rho}/\Gamma_{\pi f_2(1270)}$  are presented by making the comparison with the corresponding experimental results, where most theoretical values are consistent with the experimental data if  one considers the experimental errors. The above study supports $a_2(1700)$ as a $2^3P_2$ state in the $a_2$ meson family.

\begin{figure}[htb]
\begin{center}
\scalebox{1.05}{\includegraphics[width=\columnwidth]{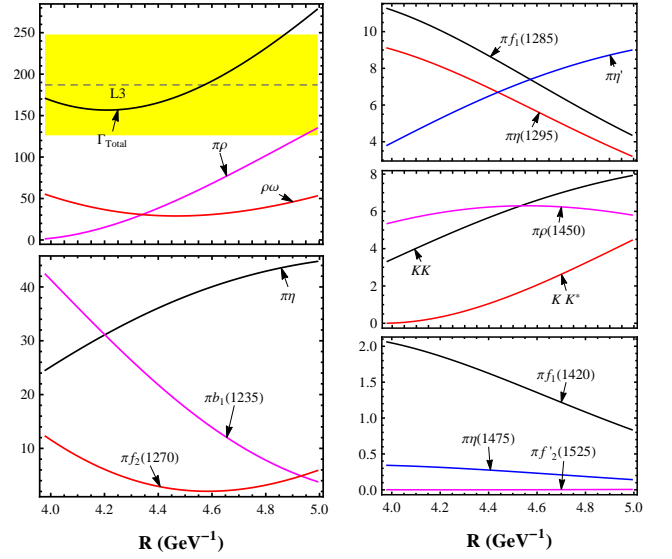}}
\caption{The dependence of the partial and total decay widths of $a_2(1700)$  as the $2^3P_2$ state on the $R$ value. Here, the dashed line with the  yellow band denotes the experimental total width given in Ref. \cite{Acciarri:2000ex}.  All results are in units of MeV.}\label{fig:1700}
\end{center}
\end{figure}

\begin{table}[htbp]
\caption{The comparison of the theoretical and experimental values for $a_2(1700)$. Here, $\Gamma_{Total}$, $\Gamma_{\pi\eta}$, $\Gamma_{KK}$ are in units of MeV.
 \label{tab:1700}}
\begin{tabular}{lcccccccc}
\toprule[1pt]
\toprule[1pt]
 {Item} &                                                  {This work} &                    {Experimental data}  \\     
 \midrule[1pt]
{$\Gamma _{Total}$} &                                            $161$ &            {$194\pm 40$~\cite{Beringer:1900zz}}  \\     

{$\Gamma _{\pi \eta}$} &                                     {$35$} &                    {$9.5\pm 2.0$}~\cite{Beringer:1900zz,Shchegelsky:2006et} \\     

{$\Gamma _{K  K}$} &                                            {$5.4$} &        {$5.0\pm 3.0$}~\cite{Beringer:1900zz,Shchegelsky:2006et} \\       

{$\Gamma _{\pi\rho}/\Gamma _{\pi f_2(1270)}$} &                 8.4       &        {$3.4\pm0.4\pm0.1$~\cite{Beringer:1900zz,Shchegelsky:2006es}} \\       

\bottomrule[1pt]
\bottomrule[1pt]
\end{tabular}
\end{table}

Along with investigating the decay behaviors of $a_2(1320)$ and $a_2(1700)$, the reliability of the QPC model can be further tested
in this work, which makes us safely apply this model to the remaining $a_2$ states.

In the following, we illustrate the decay behavior of $a_2(2030)$ as a $1^3F_2$ state, where its total and partial decay widths are shown in Fig. \ref{fig:2030}. Our calculation shows that $a_2(2030)$ as a $1^3F_2$ state has a very broad width, i.e.,
the total width can reach up to (730$\sim$830) MeV corresponding to $R=(4.00\sim 5.00)$  GeV$^{-1}$, which is not strongly dependent on the $R$ value. When comparing the calculated total width with the experimental result, we find that the theoretical result is far larger than the experimental width of $a_2(2030)$ \cite{Anisovich:2001pn}. At present, $a_2(2030)$ was only reported in Ref. \cite{Anisovich:2001pn,Shchegelsky:2006es}. Thus, we suggest further experimental study to measure the resonance parameters of $a_2(2030)$, which is important to clarify the difference between theory and experiment. From Fig. \ref{fig:2030}, we obtain that $\pi b_1(1235)$, $\rho h_1(1170)$, $\pi \eta_2(1645)$, $\eta a_1(1260)$, and $\pi f_2(1270)$ are main decay modes of $a_2(2030)$, where $\pi f_2(1270)$ was already observed in the Crystal Barrel experiments  \cite{Anisovich:2001pn,Bugg:2004xu,Bugg:2012yt}. The details of other decay information of $a_2(2030)$ can be found in Fig. \ref{fig:2030}.
The predicted decay behaviors of $a_2(2030)$ are valuable to confirm this state by future experiments.

\begin{figure*}[htbp]
\begin{center}
\scalebox{2.1}{\includegraphics[width=\columnwidth]{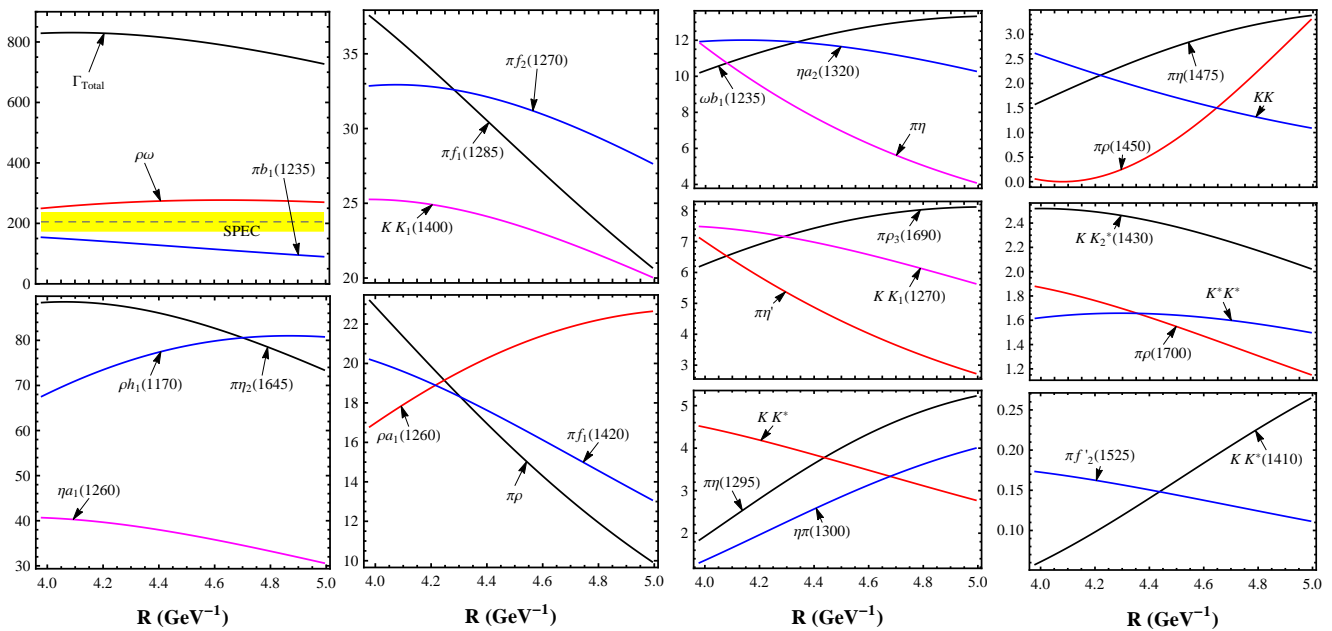}}
\caption{The variation of the partial and total decay widths of $a_2(2030)$  as a $1^3F_2$ state in the $R$ value. Here, the dashed line with the yellow band is the experimental total width from Ref. \cite{Anisovich:2001pn}.  All results are in units of MeV.}\label{fig:2030}
\end{center}
\end{figure*}

\subsubsection{The possibility of $a_2(1950)$ or $a_2(2175)$ as the $3^3P_2$ state}

As shown in Table \ref{tab:2}, there are two possible candidates for the $3^3P_2$ state, i.e., $a_2(1950)$ and $a_2(2175)$.
Thus, the study of the decay behaviors of $a_2(1950)$ and $a_2(2175)$ is helpful to distinguish these two possibilities. In Table \ref{tab:3}, the allowed decay channels of $a_2(1950)$ and $a_2(2175)$ are listed. In the following calculation, we separately discuss the decay behaviors of $a_2(1950)$ and $a_2(2175)$ under the $3^3P_2$ assignment.

In Fig. \ref{fig:1950}, we show the total and partial decay widths of
$a_2(1950)$ dependent on the $R$ value.
Here, the calculated total width overlaps with the SPEC experimental data \cite{Anisovich:2001pn} with $R=(4.73\sim 5.14)$~GeV$^{-1}$. The result of the corresponding partial decay width shows that $a_2(1950)$ dominantly decays into $ \pi\rho$, $\pi\eta $, $\rho\omega$,  and $\pi \eta^\prime$ (see Fig. \ref{fig:1950} for more details). In addition, we also select some typical ratios in Table \ref{tab:1950} that are weakly dependent on the $R$ values.

\begin{figure*}[htbp]
\begin{center}
\scalebox{2.1}{\includegraphics[width=\columnwidth]{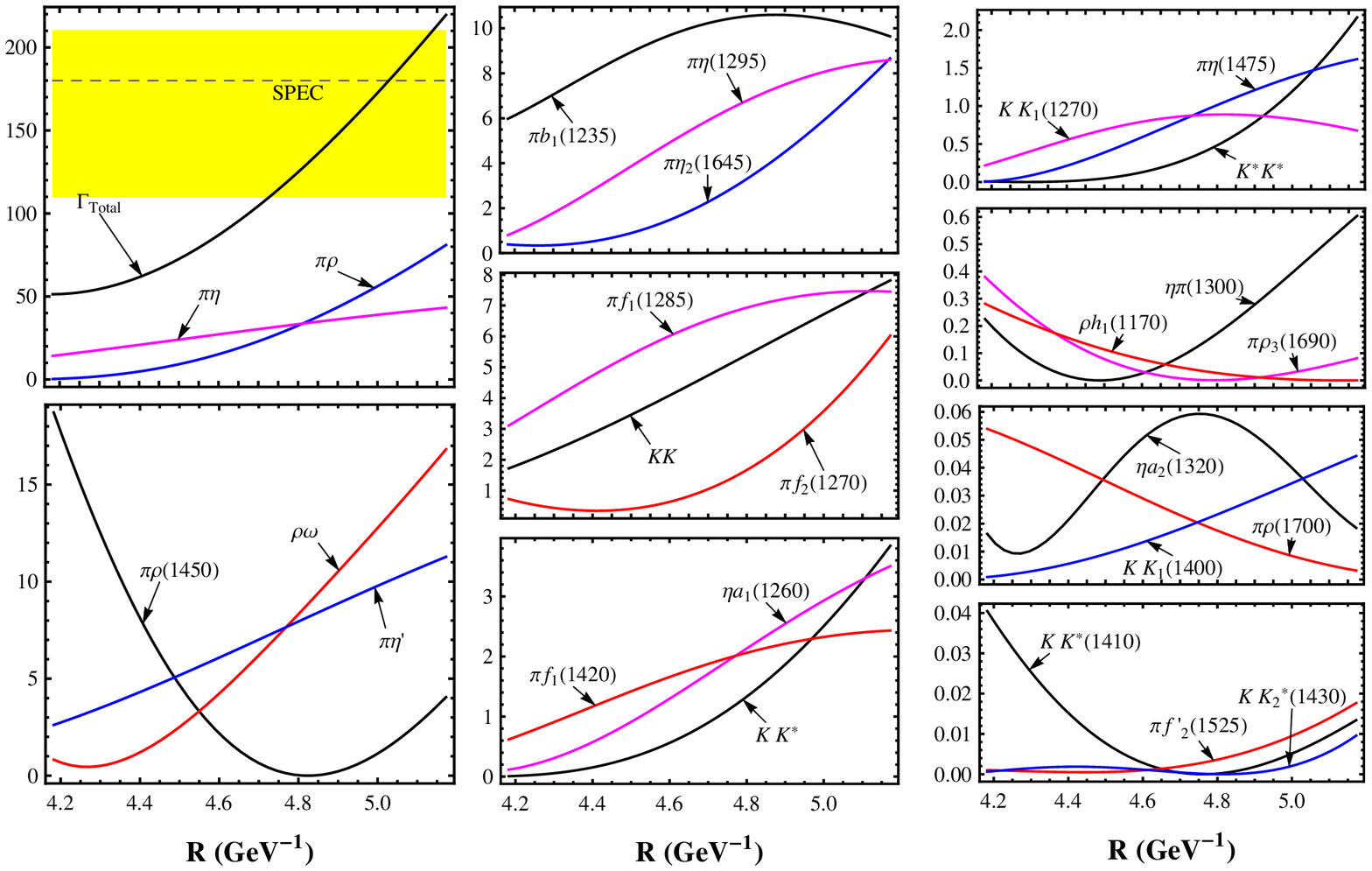}}
\caption{The dependence of the partial and total decay widths of $a_2(1950)$ as the $3^3P_2$ state on the $R$ value. Here, the dashed line with the yellow band is the experimental total width from Ref. \cite{Anisovich:2001pn}.  All results are in units of MeV.}
\label{fig:1950}
\end{center}
\end{figure*}

\begin{table}[htbp]
\tabcolsep=2pt
\caption{The typical ratios relevant to the decay behavior of $a_2(1950)$. Here, these results correspond to the range $R=(4.73\sim 5.14)$ GeV$^{-1}$.
 \label{tab:1950}}
\begin{tabular}{cccccc}
\toprule[1pt]
\toprule[1pt]
Ratios &         Value                                              &Ratios &    Value                                                  \\     
 \midrule[1pt]
$\Gamma _{\pi \rho}/\Gamma _{Total}$ & 0.229$\sim$0.364&
$\Gamma _{\pi \eta}/\Gamma _{Total}$&0.200$\sim$0.285\\
$\Gamma _{\rho \omega }/\Gamma _{Total}$&0.0613$\sim$0.0763&
$\Gamma _{\rho \omega }/\Gamma _{\pi \rho }$&0.210$\sim$0.267\\
$\Gamma _{\rho \omega }/\Gamma _{\pi \eta }$&0.215$\sim$0.381&
$\Gamma _{\text{$\pi \eta'$}}/\Gamma _{Total}$&0.0520$\sim$0.0663\\
$\Gamma _{\text{$\pi \eta'$}}/\Gamma _{\pi \eta }$&0.233$\sim$0.260&
$\Gamma _{\text{$\pi \eta'$}}/\Gamma _{\rho \omega }$&0.681$\sim$1.08\\
$\Gamma _{\pi b_1\text{(1235)}}/\Gamma _{\pi \eta }$&0.229$\sim$0.333&
$\Gamma _{\pi b_1\text{(1235)}}/\Gamma _{\text{$\pi \eta'$}}$&0.881$\sim$1.43\\
$\Gamma _{KK}/\Gamma _{Total}$&0.0360$\sim$0.0449&
$\Gamma _{KK}/\Gamma _{\pi \eta }$&0.158$\sim$0.180\\
$\Gamma _{KK}/\Gamma _{\rho \omega }$&0.472$\sim$0.732&
$\Gamma _{KK}/\Gamma _{\text{$\pi \eta'$}}$&0.676$\sim$0.692\\
$\Gamma _{KK}/\Gamma _{\text{$\pi \eta $(1295)}}$&0.796$\sim$0.897&
$\Gamma _{KK}/\Gamma _{\pi f_1\text{(1285)}}$&0.735$\sim$1.03\\
\bottomrule[1pt]
\bottomrule[1pt]
\end{tabular}
\end{table}

\begin{table}[htbp]
\tabcolsep=2pt
\caption{The typical ratios relevant to the decay behavior of $a_2(2175)$ as the 3$^3P_2$ or a 4$^3P_2$ state, where we take $R=(4.20\sim 4.72)$ and (5.46$\sim$ 5.78) GeV$^{-1}$ corresponding to the 3$^3P_2$ and 4$^3P_2$ states, respectively.
 \label{tab:21753p}}
\begin{tabular}{cccccc}
\toprule[1pt]
\toprule[1pt]
Ratios &3P state &   4P state                                               \\     
 \midrule[1pt]
$\Gamma _{\pi \rho }/\Gamma _{Total}$ &0.0483$\sim$0.289& 0.350$\sim$0.377\\
$\Gamma _{\pi \eta }/\Gamma _{Total}$&0.107$\sim$0.164&0.140$\sim$0.180\\
$\Gamma _{\pi b_1\text{(1235)}}/\Gamma _{Total}$&0.0443$\sim$0.0704&0.0345$\sim$0.0503\\
$\Gamma _{\pi b_1\text{(1235)}}/\Gamma _{\pi \rho }$&0.226$\sim$0.915&0.0885$\sim$0.144\\
$\Gamma _{\pi b_1\text{(1235)}}/\Gamma _{\pi \eta }$&0.413$\sim$0.431&0.246$\sim$0.279\\
$\Gamma _{\rho \omega }/\Gamma _{Total}$&0.0423$\sim$0.0693&0.0294$\sim$0.0396\\
$\Gamma _{\rho \omega }/\Gamma _{\pi \rho }$&0.146$\sim$1.27&0.0840$\sim$0.105\\
$\Gamma _{\rho \omega }/\Gamma _{\pi \eta }$&0.268$\sim$0.570&0.163$\sim$0.282\\
$\Gamma _{\rho \omega }/\Gamma _{\pi b_1\text{(1235)}}$&0.648$\sim$1.38&0.585$\sim$1.15\\
$\Gamma _{KK}/\Gamma _{Total}$&0.0147$\sim$0.0285&0.0283$\sim$0.0332\\
$\Gamma _{KK}/\Gamma _{\pi \rho }$&0.0984$\sim$0.304&0.0750$\sim$0.0947\\
$\Gamma _{KK}/\Gamma _{\pi \eta }$&0.137$\sim$0.180&0.184$\sim$0.202\\
$\Gamma _{KK}/\Gamma _{\pi b_1\text{(1235)}}$&0.331$\sim$0.436&0.659$\sim$0.821\\
$\Gamma _{KK}/\Gamma _{\rho \omega }$&0.240$\sim$0.673&0.715$\sim$1.14\\
$\Gamma _{KK}/\Gamma _{\text{$\pi \eta'$}}$&0.525$\sim$0.577&0.582$\sim$0.602\\
\bottomrule[1pt]
\bottomrule[1pt]
\end{tabular}
\end{table}

\begin{table*}[htbp]
\tabcolsep=2pt
\caption{The typical ratios relevant to the decay behavior of $a_2(2255)$ as the 4$^3P_2$ state. Here, these results correspond to the range $R=(5.09\sim 5.16)$ GeV$^{-1}$.
 \label{tab:22254P}}
\begin{tabular}{cccccc}
\toprule[1pt]
\toprule[1pt]
Ratios &         Value                                              &Ratios &    Value                                                  \\     
 \midrule[1pt]
$\Gamma _{\pi \rho }/\Gamma _{\text{Total}}$&0.300$\sim$0.314&
$\Gamma _{\pi \eta }/\Gamma _{\text{Total}}$&0.195$\sim$0.209\\
$\Gamma _{\pi \eta }/\Gamma _{\pi \rho }$&0.620$\sim$0.696&
$\Gamma _{\pi b_1\text{(1235)}}/\Gamma _{\text{Total}}$&0.0651$\sim$0.0690\\
$\Gamma _{\pi b_1\text{(1235)}}/\Gamma _{\pi \rho }$&0.207$\sim$0.2300&
$\Gamma _{\pi b_1\text{(1235)}}/\Gamma _{\pi \eta }$&0.330$\sim$0.334\\
$\Gamma _{\text{$\pi \eta'$}}/\Gamma _{\text{Total}}$&0.0623$\sim$0.0655&
$\Gamma _{\text{$\pi \eta'$}}/\Gamma _{\pi \rho }$&0.198$\sim$0.218\\
$\Gamma _{\text{$\pi \eta'$}}/\Gamma _{\pi \eta }$&0.314$\sim$0.320&
$\Gamma _{\text{$\pi \eta'$}}/\Gamma _{\pi b_1\text{(1235)}}$&0.949$\sim$0.957\\
$\Gamma _{\text{$\pi \eta $(1295)}}/\Gamma _{\text{Total}}$&0.0532$\sim$0.0565&
$\Gamma _{\text{$\pi \eta $(1295)}}/\Gamma _{\pi \rho }$&0.169$\sim$0.188\\
$\Gamma _{\text{$\pi \eta $(1295)}}/\Gamma _{\pi \eta }$&0.271$\sim$0.273&
$\Gamma _{\text{$\pi \eta $(1295)}}/\Gamma _{\pi b_1\text{(1235)}}$&0.817$\sim$0.819\\
$\Gamma _{\text{$\pi \eta $(1295)}}/\Gamma _{\text{$\pi \eta'$}}$&0.854$\sim$0.863&
$\Gamma _{\pi f_2\text{(1270)}}/\Gamma _{\text{Total}}$&0.0408$\sim$0.0443\\
$\Gamma _{\pi f_2\text{(1270)}}/\Gamma _{\pi \rho }$&0.136$\sim$0.141&
$\Gamma _{\pi f_2\text{(1270)}}/\Gamma _{\pi \eta }$&0.195$\sim$0.228\\
$\Gamma _{\pi f_2\text{(1270)}}/\Gamma _{\pi b_1\text{(1235)}}$&0.591$\sim$0.681&
$\Gamma _{\pi f_2\text{(1270)}}/\Gamma _{\text{$\pi \eta'$}}$&0.623$\sim$0.712\\
$\Gamma _{\pi f_2\text{(1270)}}/\Gamma _{\text{$\pi \eta $(1295)}}$&0.722$\sim$0.834&
$\Gamma _{KK}/\Gamma _{\text{Total}}$&0.0335$\sim$0.0349\\
$\Gamma _{KK}/\Gamma _{\pi \rho }$&0.107$\sim$0.116&
$\Gamma _{KK}/\Gamma _{\pi \eta }$&0.167$\sim$0.172\\
$\Gamma _{KK}/\Gamma _{\pi b_1\text{(1235)}}$&0.505$\sim$0.515&
$\Gamma _{KK}/\Gamma _{\text{$\pi \eta'$}}$&0.532$\sim$0.538\\
$\Gamma _{KK}/\Gamma _{\text{$\pi \eta $(1295)}}$&0.617$\sim$0.631&
$\Gamma _{KK}/\Gamma _{\pi f_2\text{(1270)}}$&0.756$\sim$0.855\\
$\Gamma _{\pi f_1\text{(1285)}}/\Gamma _{\text{Total}}$&0.0313$\sim$0.0342&
$\Gamma _{\pi f_1\text{(1285)}}/\Gamma _{\pi \rho }$&0.0996$\sim$0.114\\
$\Gamma _{\pi f_1\text{(1285)}}/\Gamma _{\pi \eta }$&0.161$\sim$0.164&
$\Gamma _{\pi f_1\text{(1285)}}/\Gamma _{\pi b_1\text{(1235)}}$&0.481$\sim$0.496\\
$\Gamma _{\pi f_1\text{(1285)}}/\Gamma _{\text{$\pi \eta'$}}$&0.502$\sim$0.522&
$\Gamma _{\pi f_1\text{(1285)}}/\Gamma _{\text{$\pi \eta $(1295)}}$&0.589$\sim$0.605\\
$\Gamma _{\pi f_1\text{(1285)}}/\Gamma _{\pi f_2\text{(1270)}}$&0.705$\sim$0.839&
$\Gamma _{\pi f_1\text{(1285)}}/\Gamma _{KK}$&0.933$\sim$0.982\\
$\Gamma _{\text{$\pi \rho $(1450)}}/\Gamma _{\text{Total}}$&0.0201$\sim$0.0275&
$\Gamma _{\text{$\pi \rho $(1450)}}/\Gamma _{\pi \rho }$&0.0669$\sim$0.0875\\
$\Gamma _{\text{$\pi \rho $(1450)}}/\Gamma _{\pi \eta }$&0.0962$\sim$0.141&
$\Gamma _{\text{$\pi \rho $(1450)}}/\Gamma _{\pi b_1\text{(1235)}}$&0.291$\sim$0.422\\
$\Gamma _{\text{$\pi \rho $(1450)}}/\Gamma _{\text{$\pi \eta'$}}$&0.307$\sim$0.441&
$\Gamma _{\text{$\pi \rho $(1450)}}/\Gamma _{\text{$\pi \eta $(1295)}}$&0.355$\sim$0.517\\
$\Gamma _{\text{$\pi \rho $(1450)}}/\Gamma _{\pi f_2\text{(1270)}}$&0.493$\sim$0.619&
$\Gamma _{\text{$\pi \rho $(1450)}}/\Gamma _{KK}$&0.576$\sim$0.819\\
$\Gamma _{\text{$\pi \rho $(1450)}}/\Gamma _{\pi f_1\text{(1285)}}$&0.587$\sim$0.878&
$\Gamma _{\pi \eta _2\text{(1645)}}/\Gamma _{\text{Total}}$&0.0268$\sim$0.0272\\
$\Gamma _{\pi \eta _2\text{(1645)}}/\Gamma _{\pi \rho }$&0.0868$\sim$0.0893&
$\Gamma _{\pi \eta _2\text{(1645)}}/\Gamma _{\pi \eta }$&0.128$\sim$0.140\\
$\Gamma _{\pi \eta _2\text{(1645)}}/\Gamma _{\pi b_1\text{(1235)}}$&0.388$\sim$0.419&
$\Gamma _{\pi \eta _2\text{(1645)}}/\Gamma _{\text{$\pi \eta'$}}$&0.409$\sim$0.438\\
$\Gamma _{\pi \eta _2\text{(1645)}}/\Gamma _{\text{$\pi \eta $(1295)}}$&0.474$\sim$0.513&
$\Gamma _{\pi \eta _2\text{(1645)}}/\Gamma _{\pi f_2\text{(1270)}}$&0.614$\sim$0.657\\
$\Gamma _{\pi \eta _2\text{(1645)}}/\Gamma _{KK}$&0.769$\sim$0.813&
$\Gamma _{\pi \eta _2\text{(1645)}}/\Gamma _{\pi f_1\text{(1285)}}$&0.784$\sim$0.871\\
$\Gamma _{\pi \eta _2\text{(1645)}}/\Gamma _{\text{$\pi \rho $(1450)}}$&0.992$\sim$1.33&
$\Gamma _{\rho \omega }/\Gamma _{\text{Total}}$&0.0217$\sim$0.0225\\
$\Gamma _{\rho \omega }/\Gamma _{\pi \rho }$&0.0693$\sim$0.0751&
$\Gamma _{\rho \omega }/\Gamma _{\pi \eta }$&0.108$\sim$0.112\\
$\Gamma _{\rho \omega }/\Gamma _{\pi b_1\text{(1235)}}$&0.327$\sim$0.334&
$\Gamma _{\rho \omega }/\Gamma _{\text{$\pi \eta'$}}$&0.344$\sim$0.349\\
$\Gamma _{\rho \omega }/\Gamma _{\text{$\pi \eta $(1295)}}$&0.399$\sim$0.409&
$\Gamma _{\rho \omega }/\Gamma _{\pi f_2\text{(1270)}}$&0.490$\sim$0.553\\
$\Gamma _{\rho \omega }/\Gamma _{KK}$&0.645$\sim$0.649&
$\Gamma _{\rho \omega }/\Gamma _{\pi f_1\text{(1285)}}$&0.659$\sim$0.695\\
$\Gamma _{\rho \omega }/\Gamma _{\text{$\pi \rho $(1450)}}$&0.792$\sim$1.12&
$\Gamma _{\rho \omega }/\Gamma _{\pi \eta _2\text{(1645)}}$&0.798$\sim$0.841\\
\bottomrule[1pt]
\bottomrule[1pt]
\end{tabular}
\end{table*}

If $a_2(2175)$ is the $3^3P_2$ state, the obtained total decay width can be fitted with the present experimental width of $a_2(2175)$ when taking $R=(4.20\sim 4.72)$ GeV$^{-1}$. Thus, its partial decay behavior is crucial when we want to distinguish $a_2(1950)$ and $a_2(2175)$ as the $3^3P_2$ state.
Here, we get that the main decay channels of $a_2(2175)$ are $\pi\rho$, $\rho a_2(1320)$, and $\pi\eta$. The detailed decay information of $a_2(2175)$ is collected in Fig. \ref{fig:2175}. By this study, we can see that the partial decay behavior of
$a_2(2175)$ is indeed different from that of $a_2(1950)$. In particular, there are slight differences in the corresponding typical ratios listed in Tables \ref{tab:1950} and \ref{tab:21753p}. We expect further experimental study on $a_2(1950)$ and $a_2(2175)$ to confirm these theoretical results in future.

\subsubsection{The possibility of $a_2(2175)$ or $a_2(2255)$ as the $4^3P_2$ state}

The mass spectrum analysis \cite{Anisovich:2001pn,Anisovich:2002us} shows that both $a_2(2175)$ and $a_2(2255)$ can be the candidates of the $4^3P_2$ state. In the following discussion, we combine the experimental data with the calculated result to predict the partial decay width of $a_2(2175)$ and $a_2(2255)$ under the assignment of the $4^3P_2$ state.

If $a_2(2175)$ is the $4^3P_2$ state, we find that the obtained total decay width can reproduce the experimental data in Ref. \cite{Anisovich:2001pn}, where the $R$ range is taken as $(5.46\sim5.78)$ GeV$^{-1}$. The corresponding partial decay widths are also listed in Fig. \ref{fig:21754P}, which shows that $\pi\rho$ and $\pi\eta$ are main decay modes. Further giving more abundant information to future experiments, we also collect the typical ratios weakly dependent on the $R$ value in Table \ref{tab:21753p}.

Similarly, we also get the total and partial decay widths of $a_2(2255)$ as the $4^3P_2$ state, which are shown in Fig. \ref{fig:22254P}. The main decay modes of $a_2(2255)$ as the $4^3P_2$ state are $\pi\rho$ and $\pi\eta$. In Table \ref{tab:22254P}, we predict some typical ratios, which can test the $4^3P_2$ state assignment to $a_2(2255)$ in future experiments.

\subsubsection{$a_2(2255)$ as the $2^3F_2$ state}

Finally, we need to discuss the possibility of $a_2(2225)$ as the $2^3F_2$ state by studying decay behavior.
The comparison between the calculated total decay width and the experimental one of $a_2(2255)$ is shown in Fig. \ref{fig:22252F}, which shows that the experimental data cannot be reproduced, i.e., our result is larger than the experimental value. There are two possibilities for this:

1. The $2^3F_2$ assignment to $a_2(2255)$ is not suitable. However, before definitely adopting this conclusion, a more precise measurement of the resonance parameters of $a_2(2255)$ is necessary since there is only one experiment relevant to $a_2(2255)$ \cite{Anisovich:2001pn} at present.

2. If $a_2(2255)$ is a $2^3F_2$ state, the corresponding partial decay widths are listed in Fig. \ref{fig:22252F}. Here, $\rho\omega$ and $\rho a_1(1260)$ are main decay channels. Thus, carrying out the search for these predicted main decay modes will be helpful to clarity whether $a_2(2255)$ as the $2^3F_2$ state
is suitable or not.

\section{Summary}

In this work, we have systematically calculated the two-body OZI-allowed strong decays of the observed $a_2$ states when they are categorized into the $a_2$ meson family. By comparing our results with the present experimental data, the $n^{2S+1}L_J$ assignments to the observed states can be tested. What is more important is that in this work we have predicted the partial decay widths of the $a_2$ states, which provides important and valuable information on further experimental searchs for the $a_2$ states.

As indicated in the review of the experimental status of the $a_2$ states in Sec. \ref{sec1}, the experimental data of these states are not abundant at present, especially the $a_2$ states with higher mass. Hopefully our work can inspire the experimentalist's interest in exploring the $a_2$ states. We also suggest future experiments to measure the resonance parameters of the observed $a_2$ states since these parameters are crucial to establish the $a_2$ meson family.

The BESIII experiment and the forthcoming PANDA experiment will be good platforms to carry out the experimental study of $a_2$, and we expect further experimental progress on the $a_2$ states.


\begin{figure*}[htbp]
\begin{center}
\scalebox{1.8}{\includegraphics[width=\columnwidth]{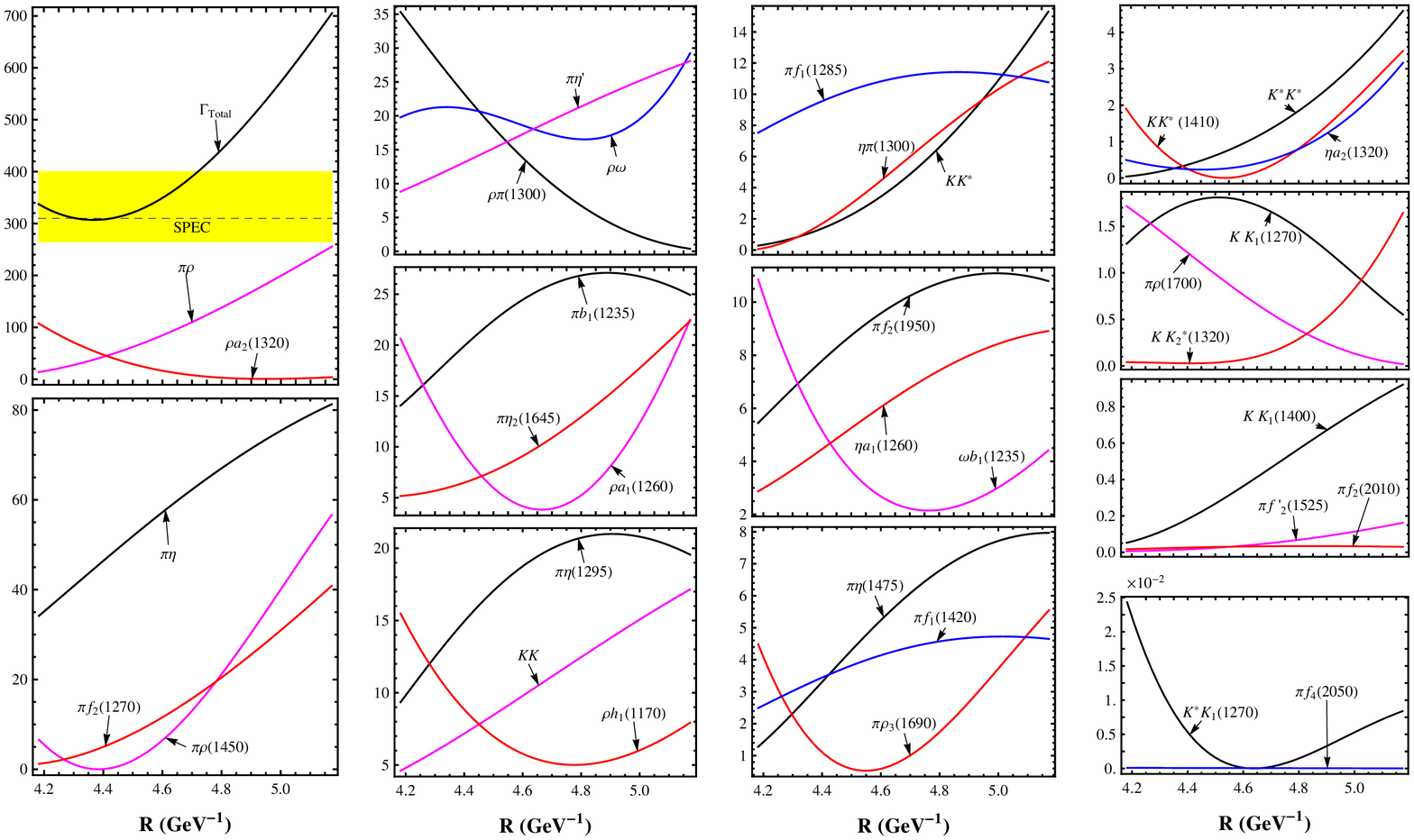}}
\caption{The decay behavior of $a_2(2175)$ as a $ 3^3P_2$ state. Here, the dashed line with the yellow band is the experimental total width given in Ref. \cite{Anisovich:2001pn}. All results are in units of MeV.}\label{fig:2175}
\end{center}
\end{figure*}

\begin{figure*}[htbp]
\begin{center}
\scalebox{1.8}{\includegraphics[width=\columnwidth]{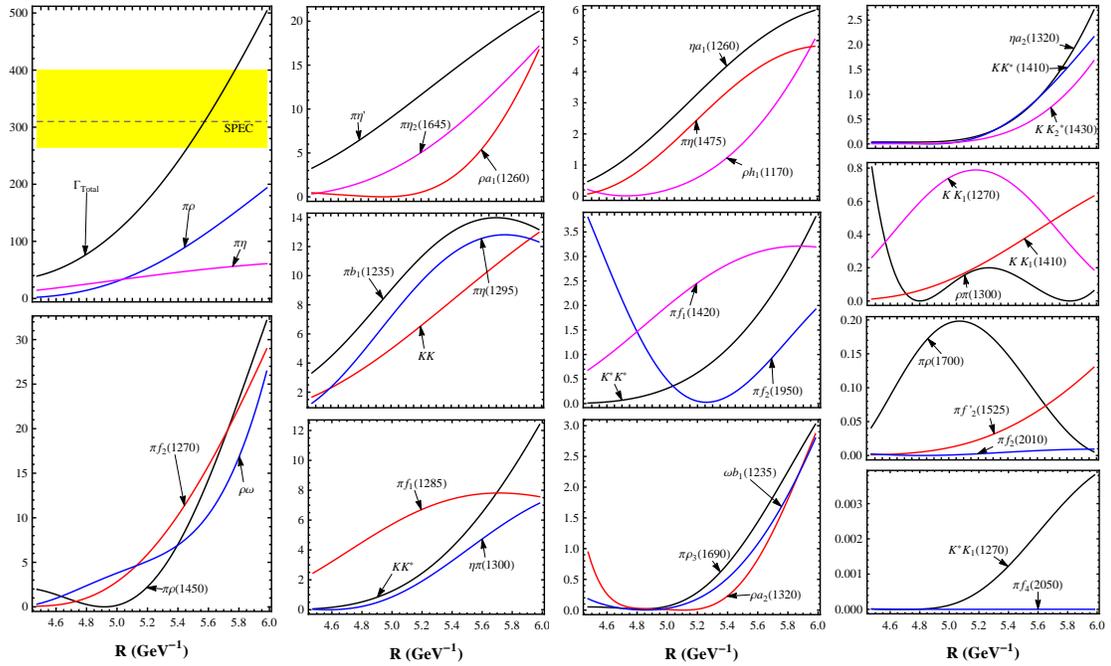}}
\caption{The obtained total and partial decay widths of $a_2(2175)$ as a $ 4^3P_2$ state on the $R$ value. Here, the dashed line with the yellow band is the experimental total width in Ref. \cite{Anisovich:2001pn}. All results are in units of MeV. }\label{fig:21754P}
\end{center}
\end{figure*}

\begin{figure*}[htbp]
\begin{center}
\scalebox{1.7}{\includegraphics[width=\columnwidth]{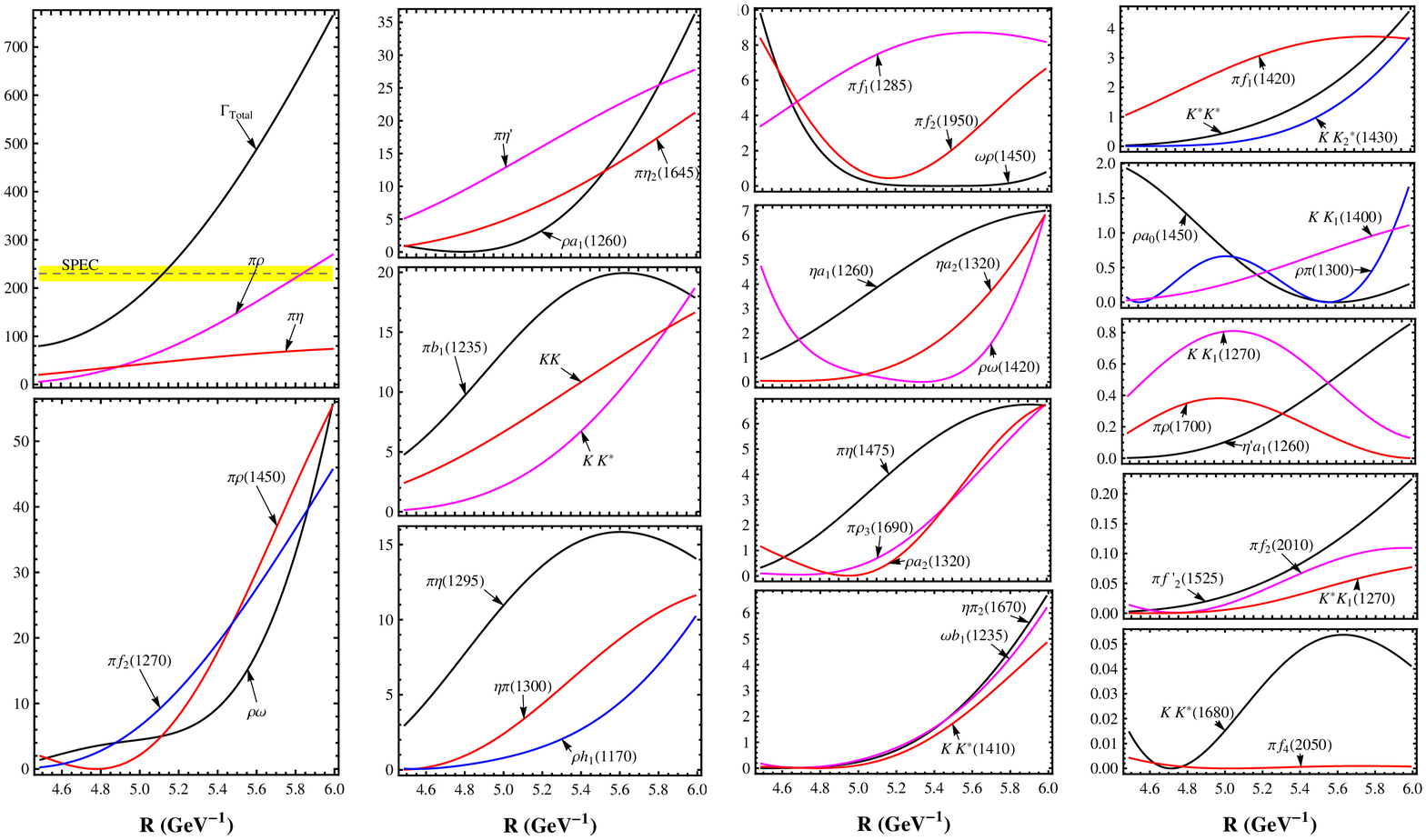}}
\caption{The total and partial decay widths of $a_2(2255)$ as a $ 4^3P_2$ state. Here, the dashed line with the band is the experimental total width from Ref. \cite{Anisovich:2001pn}. All results are in units of MeV.}\label{fig:22254P}
\end{center}
\end{figure*}

\begin{figure*}[htbp]
\begin{center}
\scalebox{1.7}{\includegraphics[width=\columnwidth]{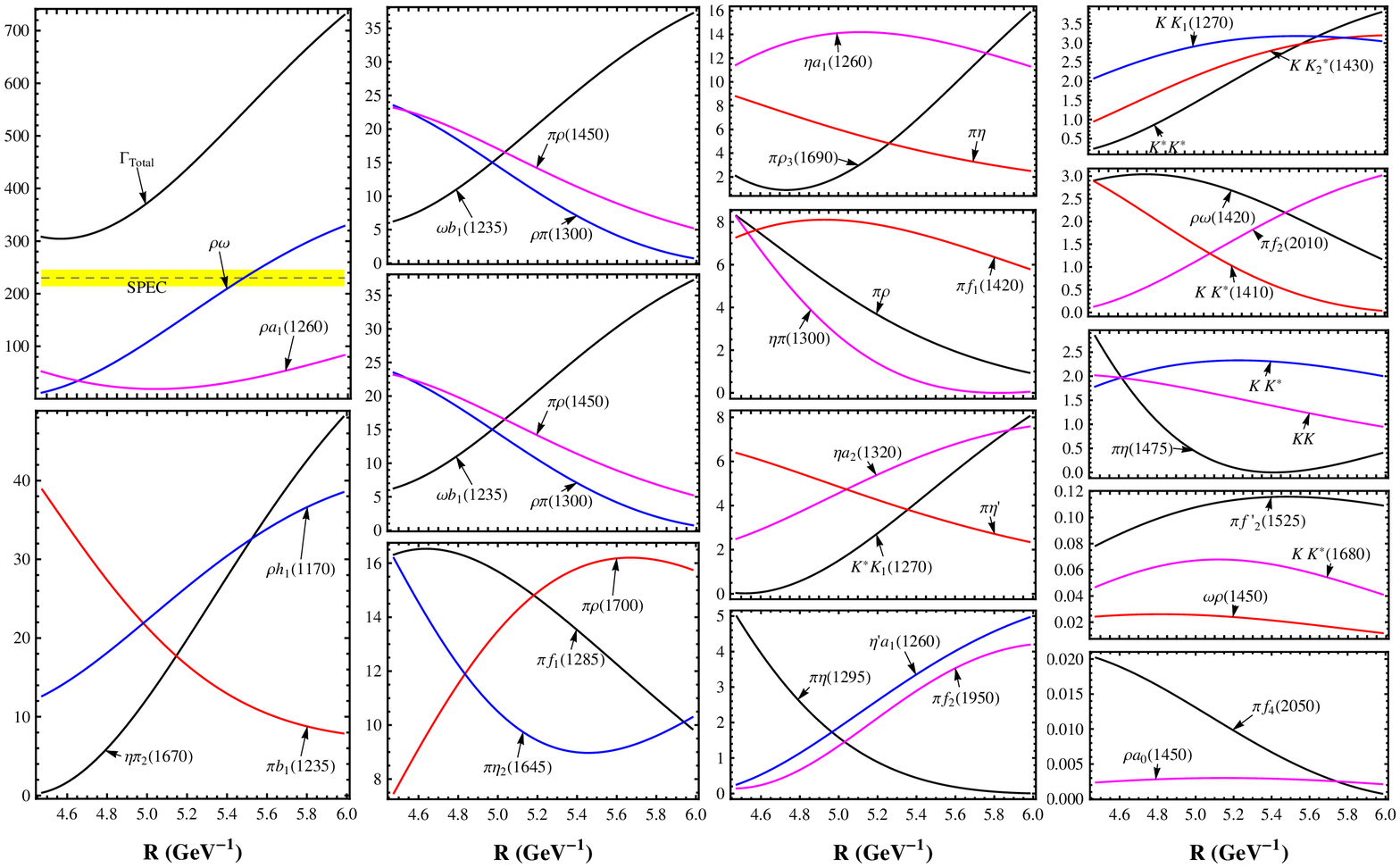}}
\caption{The variation of the total and partial decay widths of $a_2(2255)$ to the $R$ value. Here, $a_2(2255)$ is assigned to be a $ 2^3F_2$ state. The dashed line with the band is the experimental total width from Ref. \cite{Anisovich:2001pn}. All results are in units of MeV.}\label{fig:22252F}
\end{center}
\end{figure*}

\vfil

\section*{Acknowledgments}

This project is supported by the National Natural Science
Foundation of China under Grants No. 11222547, No. 11175073 and No. 11035006, the Ministry of Education of China
(SRFDP) under Grant No.
2012021111000, and the Fok Ying Tung Education Foundation
(Grant No. 131006).


\begin{thebibliography}{99}

\bibitem{Beringer:1900zz}
  J.~Beringer {\it et al.}  [Particle Data Group Collaboration],
  Phys.\ Rev.\ D {\bf 86}, 010001 (2012).

\bibitem{Anisovich:2001pn}
  A.~V.~Anisovich, C.~A.~Baker, C.~J.~Batty, D.~V.~Bugg, V.~A.~Nikonov, A.~V.~Sarantsev, V.~V.~Sarantsev and B.~S.~Zou,
  Phys.\ Lett.\ B {\bf 517}, 261 (2001)
  [arXiv:1110.0278 [hep-ex]].

\bibitem{Bugg:2004xu}
  D.~V.~Bugg,
  Phys.\ Rept.\  {\bf 397}, 257 (2004)
  [hep-ex/0412045].

\bibitem{Caso:1998tx}
  C.~Caso {\it et al.}  [Particle Data Group Collaboration],
  Eur.\ Phys.\ J.\ C {\bf 3}, 1 (1998).

\bibitem{Burakovsky:1999ug}
  L.~Burakovsky and P.~R.~Page,
  Eur.\ Phys.\ J.\ C {\bf 12}, 489 (2000)
  [hep-ph/9906282].

\bibitem{Anderson:1990mv}
  J.~D.~Anderson, M.~H.~Austern and R.~N.~Cahn,
  Phys.\ Rev.\ D {\bf 43}, 2094 (1991).

\bibitem{Ackleh:1991dy}
  E.~S.~Ackleh and T.~Barnes,
  Phys.\ Rev.\ D {\bf 45}, 232 (1992).

\bibitem{Munz:1996hb}
  C.~R.~Munz,
  Nucl.\ Phys.\ A {\bf 609}, 364 (1996)
  [hep-ph/9601206].

\bibitem{Barnes:1996ff}
  T.~Barnes, F.~E.~Close, P.~R.~Page and E.~S.~Swanson,
  Phys.\ Rev.\ D {\bf 55}, 4157 (1997)
  [hep-ph/9609339].

\bibitem{Amsler:2008zzb}
  C.~Amsler {\it et al.}  [Particle Data Group Collaboration],
  Phys.\ Lett.\ B {\bf 667}, 1 (2008).

\bibitem{Anisovich:2001pp}
  A.~V.~Anisovich, C.~A.~Baker, C.~J.~Batty, D.~V.~Bugg, V.~A.~Nikonov, A.~V.~Sarantsev, V.~V.~Sarantsev and B.~S.~Zou,
  Phys.\ Lett.\ B {\bf 517}, 273 (2001)
  [arXiv:1109.6817 [hep-ex]].

\bibitem{Anisovich:1999jv}
  A.~V.~Anisovich {\it et al.}  [Crystal Barrel Collaboration],
  Phys.\ Lett.\ B {\bf 452}, 173 (1999).

\bibitem{Hagiwara:2002fs}
  K.~Hagiwara {\it et al.}  [Particle Data Group Collaboration],
  Phys.\ Rev.\ D {\bf 66}, 010001 (2002).
  
\bibitem{Masjuan:2012gc}
  P.~Masjuan, E.~R. Arriola and W.~Broniowski,
  Phys.\ Rev.\ D {\bf 85}, 094006 (2012)
  [arXiv:1203.4782 [hep-ph]].


\bibitem{Shchegelsky:2006es}
  V.~A.~Shchegelsky, A.~V.~Sarantsev, A.~V.~Anisovich and M.~P.~Levchenko,
  Eur.\ Phys.\ J.\ A {\bf 27}, 199 (2006).

\bibitem{Lu:2004yn}
  M.~Lu {\it et al.}  [E852 Collaboration],
  Phys.\ Rev.\ Lett.\  {\bf 94}, 032002 (2005)
  [hep-ex/0405044].

\bibitem{Micu:1968mk}
  L.~Micu,
  Nucl.\ Phys.\ B {\bf 10}, 521 (1969).
  
\bibitem{Anisovich:1999jx}
  A.~V.~Anisovich {\it et al.}  [Crystal Barrel Collaboration],
  Phys.\ Lett.\ B {\bf 452}, 187 (1999).

\bibitem{Anisovich:2000kxa}
  A.~V.~Anisovich, V.~V.~Anisovich and A.~V.~Sarantsev,
  Phys.\ Rev.\ D {\bf 62}, 051502 (2000)
  [hep-ph/0003113].

\bibitem{Bugg:2012yt}
  D.~V.~Bugg,
  Phys.\ Rev.\ D {\bf 87}, no. 11, 118501 (2013)
  [arXiv:1209.3481 [hep-ph]].

\bibitem{Anisovich:2002us}
  V.~V.~Anisovich,
  Usp.\ Fiz.\ Nauk\ {\bf 47}, 49 (2004)
  [Sov. Phys.\ Usp.\  {\bf 47}, 45 (2004)]
  [hep-ph/0208123].

\bibitem{Anisovich:2003tm}
  V.~V.~Anisovich,
  AIP Conf.\ Proc.\  {\bf 717}, 441 (2004)
  [hep-ph/0310165].


  
 
\bibitem{Le Yaouanc:1972ae}
  A.~Le Yaouanc, L.~Oliver, O.~Pene and J.~C.~Raynal,
  Phys.\ Rev.\ D {\bf 8}, 2223 (1973).



\bibitem{Le Yaouanc:1973xz}
  A.~Le Yaouanc, L.~Oliver, O.~Pene and J.~-C.~Raynal,
  Phys.\ Rev.\ D {\bf 9}, 1415 (1974).

\bibitem{Le Yaouanc:1974mr}
  A.~Le Yaouanc, L.~Oliver, O.~Pene and J.~C.~Raynal,
  Phys.\ Rev.\ D {\bf 11}, 1272 (1975).

\bibitem{Le Yaouanc:1977ux}
  A.~Le Yaouanc, L.~Oliver, O.~Pene and J.~-C.~Raynal,
  Phys.\ Lett.\ B {\bf 71}, 397 (1977).

\bibitem{LeYaouanc:1977gm}
  A.~Le Yaouanc, L.~Oliver, O.~Pene and J.~C.~Raynal,
  Phys.\ Lett.\ B {\bf 72}, 57 (1977).

\bibitem{vanBeveren:1982qb}
  E.~van Beveren, G.~Rupp, T.~A.~Rijken and C.~Dullemond,
  Phys.\ Rev.\ D {\bf 27}, 1527 (1983).

\bibitem{Capstick:1993kb}
  S.~Capstick and W.~Roberts,
  Phys.\ Rev.\ D {\bf 49}, 4570 (1994)
  [nucl-th/9310030].

\bibitem{Blundell:1995ev}
  H.~G.~Blundell and S.~Godfrey,
  Phys.\ Rev.\ D {\bf 53}, 3700 (1996)
  [hep-ph/9508264].

\bibitem{Ackleh:1996yt}
  E.~S.~Ackleh, T.~Barnes and E.~S.~Swanson,
  Phys.\ Rev.\ D {\bf 54}, 6811 (1996)
  [hep-ph/9604355].

\bibitem{Capstick:1996ib}
  S.~Capstick and B.~D.~Keister,
  [nucl-th/9611055].

\bibitem{Close:2005se}
  F.~E.~Close and E.~S.~Swanson,
  Phys.\ Rev.\ D {\bf 72}, 094004 (2005)
  [hep-ph/0505206].

\bibitem{Zhang:2006yj}
  B.~Zhang, X.~Liu, W.~-Z.~Deng and S.~-L.~Zhu,
  Eur.\ Phys.\ J.\ C {\bf 50}, 617 (2007)
  [hep-ph/0609013].

\bibitem{Lu:2006ry}
  J.~Lu,  W.~-Z.~Deng, X.~-L.~Chen, and S.~-L.~Zhu,
  Phys.\ Rev.\ D {\bf 73}, 054012 (2006)
  [hep-ph/0602167].

\bibitem{Sun:2009tg}
  Z.~-F.~Sun and X.~Liu,
  Phys.\ Rev.\ D {\bf 80}, 074037 (2009)
  [arXiv:0909.1658 [hep-ph]].

\bibitem{Liu:2009fe}
  X.~Liu, Z.~-G.~Luo and Z.~-F.~Sun,
  Phys.\ Rev.\ Lett.\  {\bf 104}, 122001 (2010)
  [arXiv:0911.3694 [hep-ph]].

\bibitem{Sun:2010pg}
  Z.~-F.~Sun, J.~-S.~Yu, X.~Liu and T.~Matsuki,
  Phys.\ Rev.\ D {\bf 82}, 111501 (2010)
  [arXiv:1008.3120 [hep-ph]].

\bibitem{Rijken:2010zza}
  T.~A.~Rijken, M.~M.~Nagels and Y.~Yamamoto,
  Nucl.\ Phys.\ A {\bf 835}, 160 (2010).

\bibitem{Yu:2011ta}
  J.~-S.~Yu, Z.~-F.~Sun, X.~Liu and Q.~Zhao,
  Phys.\ Rev.\ D {\bf 83}, 114007 (2011)
  [arXiv:1104.3064 [hep-ph]].

\bibitem{Zhou:2011sp}
  Z.~-Y.~Zhou and Z.~Xiao,
  Phys.\ Rev.\ D {\bf 84}, 034023 (2011)
  [arXiv:1105.6025 [hep-ph]].

\bibitem{Wang:2012wa}
  X.~Wang, Z.~-F.~Sun, D.~-Y.~Chen, X.~Liu and T.~Matsuki,
  Phys.\ Rev.\ D {\bf 85}, 074024 (2012)
  [arXiv:1202.4139 [hep-ph]].

\bibitem{Ye:2012gu}
  Z.~-C.~Ye, X.~Wang, X.~Liu and Q.~Zhao,
  Phys.\ Rev.\ D {\bf 86}, 054025 (2012)
  [arXiv:1206.0097 [hep-ph]].

\bibitem{Sun:2013qca}
  Y.~Sun, X.~Liu and T.~Matsuki,
  Phys.\ Rev.\ D {\bf 88}, 094020 (2013)
  [arXiv:1309.2203 [hep-ph]].

\bibitem{He:2013ttg}
  L.~-P.~He, X.~Wang and X.~Liu,
  Phys.\ Rev.\ D {\bf 88}, no. 3, 034008 (2013)
  [arXiv:1306.5562 [hep-ph]].

\bibitem{Sun:2014wea}
  Y.~Sun, Q.~-T.~Song, D.~-Y.~Chen, X.~Liu and S.~-L.~Zhu,
  Phys.\ Rev.\ D {\bf 89}, 054026 (2014)
  [arXiv:1401.1595 [hep-ph]].

\bibitem{Roberts:1992js}
  W.~Roberts and B.~Silvestre- Brac,
  Acta Phys.\ Austriaca {\bf 11}, 171 (1992).

\bibitem{Blundell:1996as}
  H.~G.~Blundell,
  hep-ph/9608473.

\bibitem{Jacob:1959at}
  M.~Jacob and G.~C.~Wick,
  Annals Phys.\  {\bf 7}, 404 (1959)
  [Annals Phys.\  {\bf 281}, 774 (2000)].

\bibitem{vanBeveren:1982sj} 
  E.~van Beveren,
  Z.\ Phys.\ C {\bf 17}, 135 (1983)
  [hep-ph/0602248].
  
\bibitem{Thompson:1997bs}
  D.~R.~Thompson {\it et al.}  [E852 Collaboration],
  Phys.\ Rev.\ Lett.\  {\bf 79}, 1630 (1997)
  [hep-ex/9705011].

\bibitem{Barnham:1971nw}
  K.~W.~J.~Barnham, G.~S.~Abrams, W.~R.~Butler, D.~G.~Coyne, G.~Goldhaber, B.~H.~Hall, J.~Macnaughton and G.~H.~Trilling,
  Phys.\ Rev.\ Lett.\  {\bf 26}, 1494 (1971).

\bibitem{Bertin:1998sb}
  A.~Bertin {\it et al.}  [OBELIX Collaboration],
  Phys.\ Lett.\ B {\bf 434}, 180 (1998).

\bibitem{Abele:1997dz}
  A.~Abele {\it et al.}  [Crystal Barrel Collaboration],
  Phys.\ Lett.\ B {\bf 404}, 179 (1997).

\bibitem{Shchegelsky:2006et}
  V.~A.~Shchegelsky, A.~V.~Sarantsev, V.~A.~Nikonov and A.~V.~Anisovich,
  Eur.\ Phys.\ J.\ A {\bf 27}, 207 (2006).

\bibitem{Acciarri:2000ex}
  M.~Acciarri {\it et al.}  [L3 Collaboration],
  Phys.\ Lett.\ B {\bf 501}, 173 (2001)
  [hep-ex/0011037].

\end{thebibliography}
\end{document}